# An Angle Independent Depth Aware Fusion Beamforming Approach for Ultrafast Ultrasound Flow Imaging*

Madhavanunni A N, *Student Member, IEEE*, Mahesh Raveendranatha Panicker, *Senior Member, IEEE*

*Abstract*— In the case of vector flow imaging systems, the most employed flow estimation techniques are the directional beamforming based cross correlation and the triangulation-based autocorrelation. However, the directional beamforming-based techniques require an additional angle estimator and are not reliable if the flow angle is not constant throughout the region of interest. On the other hand, estimates with triangulation-based techniques are prone to large bias and variance at low imaging depths due to limited angle for left and right apertures. In view of this, a novel angle independent depth aware fusion beamforming approach is proposed and evaluated in this paper. The hypothesis behind the proposed approach is that the peripheral flows are transverse in nature, where directional beamforming can be employed without the need of an angle estimator and the deeper flows being non-transverse and directional, triangulation-based vector flow imaging can be employed. In the simulation study, an overall 67.62% and 74.71% reduction in magnitude bias along with a slight reduction in the standard deviation are observed with the proposed fusion beamforming approach when compared to triangulation-based beamforming and directional beamforming, respectively, when implemented individually. The efficacy of the proposed approach is demonstrated with *in-vivo* experiments.

*Clinical Relevance*— A novel fusion beamforming technique is presented that enables simultaneous visualization and diagnosis of multiple blood vessels irrespective of the depth and blood vessel orientation.

## I. INTRODUCTION

Ultrasound Doppler technique is the safest and most used non-invasive technique for blood flow imaging and has a great diagnostic value in investigating the vascular hemodynamics [1]. The advantage of being portable, repeatable and ease of imaging makes ultrasound modality to be the best choice for point of care devices in emergency medicine, ambulatory scenarios, and bedside applications. However, the evaluation of peripheral vascular system with conventional Doppler techniques are operator dependent as the velocity estimates depends on the beam to flow angle commonly known as insonification angle. This has been addressed with the vector flow imaging (VFI) techniques such as cross-beam Doppler [2], [3], vector triangulation [4], cross-correlation methods [5], [6] and multi-transmit multi-receive schemes [7]. High frame rate Doppler (ultrafast Doppler) based VFI techniques enables dynamic visualization of complex blood flow and provides an absolute quantification of the flow rate with high temporal resolution as well [7]-[9].

Typically, an ultrafast VFI system acquires ultrasound signals multiple times at very high frame rates, beamforms the received signals and performs either autocorrelation or cross-correlation of the beamformed signals to estimate the velocity. An autocorrelation based velocity estimator determines the shift in the phase of beamformed signal as a function of time as in [4], [10] whereas a cross-correlation based estimator determines the displacement of the scatterer between the emissions to compute the velocity vectors as in [11]. Among the various beamforming techniques reported in the literature, delay and sum (DAS) is the most preferred technique due to its inherent low complexity in real-time implementation despite limited resolution, contrast and large variance in the velocity estimate. There have been various attempts like directional beamforming [12], [13], multi-transmit multi-receive scheme [7] and single-transmit multi-receive scheme [14] to reduce the estimate variance in VFI.

A cross-correlation based velocity estimation scheme proposed in [12] employs a DAS based directional beamforming that focus the received signals along the direction of the flow for better accuracy. But this requires the knowledge of flow direction before beamforming and hence an additional angle estimation algorithm must be employed. In this regard, correlation and numerical triangulation-based techniques have been developed to estimate the flow angle [13], [15]. Once the flow angle is known, the coordinate system for the image is rotated and translated and the received beams are aligned in the flow direction. However, synthesizing the directional signals by rotating the grid would be difficult, especially, in the case of complex flows, where, the beam to flow angle changes throughout the region of interest (ROI) [13]. An angle independent VFI system typically adopts multi-transmit multi-receive scheme as in [7] or triangulation-based techniques as in [4] and [14]. These techniques employ autocorrelation to estimate the phase shift as a function of time to compute the velocity vectors. However, the accuracy of the velocity estimates with the above methods are dependent on the imaging depth and are poor in near field due to limited angle availability. To address the depth dependency, this paper demonstrates a novel depth aware beamforming approach by combining two VFI techniques and utilize the advantages of both the techniques.

The rest of the paper is organized as follows. The following section (Section 2) reviews the depth dependency of the state-of-the-art VFI techniques to underline the significance of a fusion beamforming approach. Further, the

*Research supported by Department of Science and Technology - Science and Engineering Research Board (DST-SERB (ECR/2018/001746)) and the Ministry of Human Resource Development (MHRD), India

Madhavanunni A N and Mahesh Raveendranatha Panicker are with Center for Computational Imaging and the Department of Electrical Engineering, Indian Institute of Technology Palakkad, India (corresponding author e-mail: 121813001@smail.iitpkd.ac.in; mahesh@iitpkd.ac.in).

proposed novel depth aware beamforming approach is discussed in detail. Section 3 discusses the performance evaluation of the proposed approach with simulation and experimental investigations. Section 4 concludes the article and discuss some of the possibilities for the future work.

*A. Depth dependency in triangulation for VFI*

The conventional single-transmit two-receive (STTR) triangulation technique adopts synthetic aperture focused insonification scheme. It uses one transmit element and two receive elements placed at either side of the transmitter, typically called as left and right (L-R) receive apertures [4]. It employs an autocorrelation-based velocity estimator that estimates the local phase and frequency of the L-R signals ($f_L$ and $f_R$) to compute the velocity components using (1) and (2). We have demonstrated an advanced triangulation technique with plane wave insonification referred as single transmit dual angle multi-receive (STDMR) scheme for transverse flow imaging in [14]. Unlike STTR, the received echoes in STDMR scheme are steered at multiple angles and DAS beamformed with dynamic receive focusing before performing the traditional autocorrelation to estimate the velocity using (1) and (2) [4], [14].

$$v \cos\theta = \frac{f_L - f_R}{\sin\alpha} \times \frac{c}{2f_0} \quad (1)$$

$$v \sin\theta = \frac{f_L + f_R}{1 + \cos\alpha} \times \frac{c}{2f_0} \quad (2)$$

where, $v\cos\theta$ and $v\sin\theta$ are the orthogonal components of the flow velocity, $f_0$ is the transmit central frequency, $c$ is the speed of sound and $\alpha$ is the transmit-receive (Tx-Rx) angle. The STDMR scheme makes a best fit out of the estimates with different $\alpha$ values so that the estimate variance is reduced. However, the choice of $\alpha$ values at low depths are considerably less because of the smaller size of the receive aperture at lower depths. This would provide significant bias in the velocity estimate at low imaging depth or for superficial flow imaging with the STDMR scheme as in [14]. This depth dependency in beamforming provides the motivation towards the idea of depth aware beamforming for VFI. To the best of our knowledge, there have been hardly any attempts towards combining two techniques for depth dependent beamforming for VFI. Exploiting the fact that the vessels closer to the skin surface are inherently transverse when compared to deeper vessels, we propose a novel fusion beamforming technique.

## II. PROPOSED DEPTH AWARE FUSION BEAMFORMING APPROACH

The proposed fusion beamforming approach employs two different techniques to beamform the received signals and estimate the velocity vectors depending on the imaging depth as illustrated in Fig. 1. It employs directional beamforming based cross correlation technique for imaging depth less than the limiting depth, $Z_L$, and triangulation based STDMR scheme is employed for imaging depth greater than $Z_L$. The $Z_L$ is defined as a hyperparameter whose value is given by (3):

$$Z_L = F_\# \times Receive\ aperture\ size \quad (3)$$

where, $F_\#$ is the F-number. For STDMR approach in [14], it has been observed that a receive aperture consisting of at least

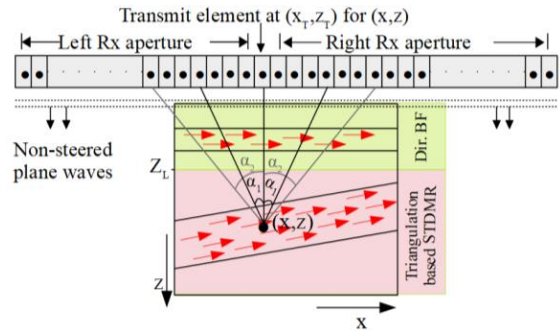

Figure 1. Illustration of the proposed depth aware fusion beamforming approach. $Z_L$ denotes the limiting depth and Dir. BF represents directional beamforming.

30 elements is required for a sufficient choice of $\alpha$ values (at least 4 or 5 different $\alpha$ values) to obtain a reliable velocity estimate. Hence a receive aperture size corresponding to a 30-element aperture is used in (3) to determine the limiting depth. It should be noted that, since the peripheral flows are inherently transverse in nature, the additional requirement for an angle estimator in directional beamforming is eliminated. Hence the received signals corresponding to the subsequent emissions are DAS beamformed and spatially cross correlated to estimate the displacement of the scatterer and flow velocity. Thus, the directional beamforming employed in the proposed approach is lot simpler than that in [13], [15], [16].

For any depth larger than $Z_L$, a triangulation based STDMR scheme is employed in which the received signals are steered at multiple angles and DAS beamformed with dynamic receive focusing. The directional beam focusing based dual apodization technique as in [17] is employed in the beamformer to ensure uniform beamwidth and better spatial resolution at larger imaging depths. The beamformed signals are then autocorrelated to estimate the phase shift and local frequency to compute the velocity components using (1) and (2). The proposed fusion approach is evaluated with exhaustive simulation study and its efficacy is demonstrated *in-vivo* as discussed in the following section.

## III. RESULTS AND DISCUSSION

*A. Simulation studies and results*

A flow phantom having multiple parabolic flows at different flow direction is simulated using Field II program [19], [20] in MATLAB R2019b. The simulation parameters closely follow the parameters chosen in [12] and is given in Table I. The simulation study employs non-steered plane wave insonification at high frame rates and considers an ROI having a transverse length of 5 mm for velocity estimation. The value of $Z_L$ is obtained as 15 mm according to (3).

For a comprehensive evaluation, the triangulation based STDMR scheme, and the directional beamforming technique have been implemented individually and their results are compared with that of the fusion approach. Ten ensembles, each consisting of 16 frames of data are used for the evaluation. Table. II presents a quantitative performance comparison for a phantom having a superficial transverse vessel and an inclined vessel (10° to the horizontal). The mean bias and the standard deviation (Std. Dev.) from the mean estimated profile are indicated as the percentage of the peak velocity. From the qualitative comparison of the results shown in Fig. 2, it is evident that the triangulation-based estimates are erroneous in most part of Vessel-1 while it approaches the true profile beyond the depth of 12 mm.

TABLE I. SIMULATION PARAMETERS

| Parameters | Symbol | Value |
|---|---|---|
| Transmit center frequency | $f_0$ | 8 MHz |
| Speed of sound | $c$ | 1540 m/s |
| Wavelength | $\lambda$ | 0.1925 mm |
| Pitch of Transmit element | $p$ | 0.1925 mm |
| Kerf | $k_e$ | 0.01 mm |
| Height of Transmit element | $h$ | 5 mm |
| No. of active elements | $N$ | 128 |
| Sampling frequency | $f_s$ | 100 MHz |
| Pulse repetition frequency | $f_{prf}$ | 15.6 kHz |
| No. of frames used | $K$ | 16 |
| Radius of vessels | $R$ | 4 mm, 5 mm |
| Peak velocities of flow | $v_0$ | 0.5 m/s, 0.5 m/s |
| Receive beamforming parameters (for both simulations and experiments) | | |
| Directional beamforming and cross correlation | | |
| F-number | $F_\#$ | 2 |
| Receive apodization | $W$ | Hanning |
| Triangulation based STDMR and autocorrelation: | | |
| F-number | $F_\#$ | 2 |
| Receive apodization | $W$ | Hanning |
| Transmit - Receive angles | $\alpha$ | 6º, 9º, 12º and 15º |

TABLE II. QUANTITATIVE COMPARISON OF VELOCITY ESTIMATES FOR SIMULATION STUDIES.

| | % Bias | % Std. Dev. |
|---|---|---|
| Dir. cross correlation | 21.46 % | 06.26 % |
| STDMR triangulation | 16.74 % | 06.98 % |
| **Fusion beamforming** | **05.42 %** | **06.24 %** |

### B. Experimental investigation and results

The proposed fusion beamforming approach is further evaluated with an *in-vivo* carotid dataset. The data is acquired using the Verasonics Vantage research ultrasound system at 10000 frames per second with a standard 128 element linear array probe (L11-5v) having a center frequency of 7.6 MHz. The *in-vivo* carotid data is acquired from a healthy volunteer by following the principles outlined in the Helsinki Declaration of 1975, as revised in 2000. The flow dynamics in the internal and external carotid artery (ICA and ECA) is evaluated and a qualitative comparison of the results obtained for *in-vivo* carotid dataset is shown in Fig. 3. Ten ensembles, each consisting of 32 frames of data are used for the evaluation. A truncated singular value decomposition filter available in the MUST toolbox [21], [22] is used as the clutter filter. The F-number is chosen as 1.71 as per [18] and a limiting depth of 15.4 mm is obtained for experimental studies as per (3). Hence, the proposed fusion beamforming chooses directional cross correlation for evaluating the ECA and triangulation based STDMR scheme for evaluating the ICA in Fig. 3(a).

A remarkable improvement in the vector flow images can be observed with the proposed fusion approach as in Fig. 3. The estimates in the ICA have a significant bias and a standard deviation of 0.896 m/s within the estimates when evaluated with directional cross correlation while a standard deviation of 0.21 m/s is observed within the estimates for ECA with triangulation. On the other hand, the standard deviation in the velocity estimates for ECA with directional cross correlation and for ICA with triangulation are observed as 0.1 m/s and 0.22 m/s, respectively, which is the same when evaluated with the proposed fusion approach.

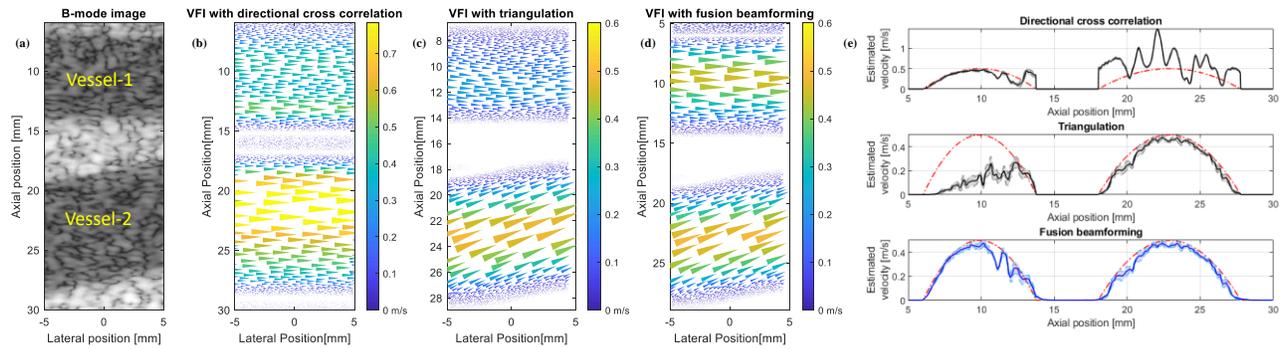

Figure 2. Simulation results: (a) B-mode image of the ROI, (b) VFI obtained with directional cross correlation, (c) VFI obtained with triangulation based STDMR scheme, (d) VFI obtained with the proposed depth aware fusion beamforming method, (e) Comparison of mean estimated velocity plots. Shaded region shows one standard deviation. Vector flow images are prepared with the help of *vplot* in MUST toolbox available online [22]. The *vplot* randomly distributes the velocity vectors as colored wedges whose areas are proportional to the velocity amplitudes. Hence the number of wedges and resolution of these VFI plots is automatically scaled by the *vplot*.

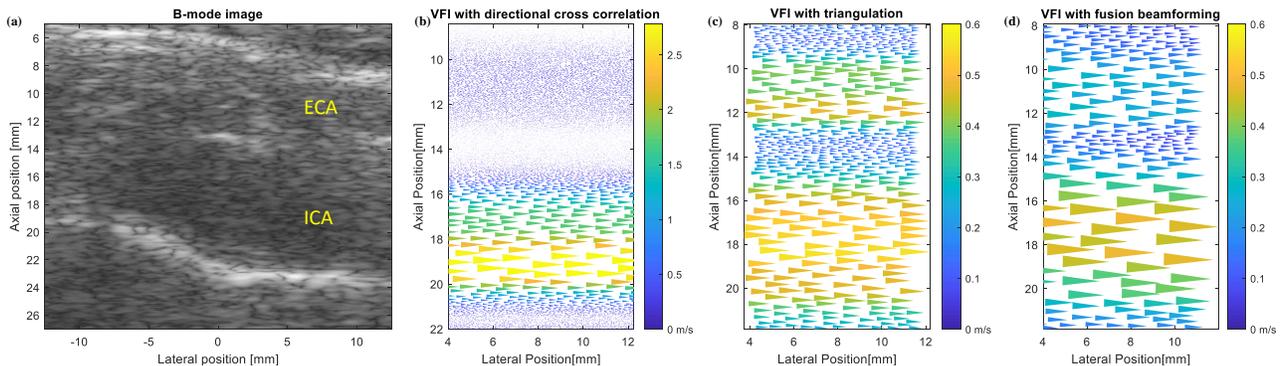

Figure 3. Experimental results for *in-vivo* carotid bifurcation: (a) B-mode image of the ROI, (b) VFI obtained with directional cross correlation, (c) VFI obtained with triangulation based STDMR scheme, (d) VFI obtained with the proposed depth aware fusion beamforming method. Vector flow images are prepared with the help of *vplot* in MUST toolbox available online [22]. The *vplot* randomly distributes the velocity vectors as colored wedges whose areas are proportional to the velocity amplitudes. Hence the number of wedges and resolution of these VFI plots is automatically scaled by the *vplot*.

It could be observed that the proposed approach provides the best estimate irrespective of the imaging depth as compared to the other two techniques when implemented individually for simulations as well as for the *in-vivo* carotid dataset. In fact, the simultaneous visualization of multi-directional flows at different depths is made possible with the proposed approach which is otherwise challenging for most of the state-of-the-art VFI techniques. Also, it is worth noting that the proposed fusion approach is an angle independent method and assumes transverse flows for superficial flow imaging.

## IV. Conclusion

This paper addresses the depth dependencies in triangulation based VFI technique and the practical challenges of directional beamforming by proposing a novel depth aware beamforming scheme with a clever fusion of both the techniques. It selects the beamforming technique according to the hyperparameter, $Z_L$, whose value is determined by the depth and the receive aperture size. It employs directional beamforming with cross correlation technique for low depth flows (depth $< Z_L$) or superficial flow imaging whereas triangulation-based autocorrelation technique is adopted for imaging the flow at larger depths (depth $\geq Z_L$). The proposed fusion approach is evaluated with an extensive simulation study and its efficacy is demonstrated with *in-vivo* experiments. The overall performance of the proposed fusion beamforming technique is observed to be superior to that of directional beamforming and triangulation-based technique when implemented individually. The fact that most of the superficial flows in the human vasculature are inherently transverse and the deep vessels are mostly inclined, justifies the choice of the techniques adopted in the proposed fusion approach. Also, it permits the user to choose the best beamforming parameters for different depths which improves the overall accuracy of the velocity estimate. However, accurate imaging of the superficial inclined micro-vessels poses a challenge and needs to be addressed. More studies are required to explore the capability of the depth dependent beamforming technique for tissue velocity imaging and elastography applications.

## Acknowledgment

The authors would like to acknowledge the funding from the Department of Science and Technology - Science and Engineering Research Board (DST-SERB (ECR/2018/001746)) and the Ministry of Human Resource Development (MHRD), India.